\documentclass[letterpaper]{article} 
\usepackage[submission]{aaai24}  
\usepackage{times}  
\usepackage{helvet}  
\usepackage{courier}  
\usepackage[hyphens]{url}  
\usepackage{graphicx} 
\usepackage{natbib}  
\usepackage{caption} 
\usepackage{mathbbol}
\usepackage{dutchcal}
\usepackage{algorithm}
\usepackage{algorithmic}
\usepackage{multirow}
\usepackage{multicol}

\title{LKPNR: \underline{L}LM and \underline{K}G for \underline{P}ersonalized \underline{N}ews \underline{R}ecommendation Framework}

\author{
    Chen Hao\textsuperscript{\rm 1}\equalcontrib,
    Xie Runfeng\textsuperscript{\rm 1}\equalcontrib,
    Cui Xiangyang\textsuperscript{\rm 1},
    Yan zhou\textsuperscript{\rm 1},
    Wang xin\textsuperscript{\rm 1},
    Xuanzhanwei\textsuperscript{\rm 1}\thanks{Corresponding author},
    Zhang kai\textsuperscript{\rm 1}\thanks{Corresponding author}\\
}

\affiliations{

\textsuperscript{\rm 1}State Key Laboratory of Communication Content Cognition, People's Daily Online, Beijing 100733, China

}

\usepackage{bibentry}

\begin{document}

\maketitle

\begin{abstract}
Accurately recommending candidate news articles to users is a basic challenge faced by personalized news recommendation systems. Traditional methods are usually difficult to grasp the complex semantic information in news texts, resulting in unsatisfactory recommendation results. Besides, these traditional methods are more friendly to active users with rich historical behaviors. However, they can not effectively solve the ``long tail problem" of inactive users. To address these issues, this research presents a novel general framework that combines Large Language Models (LLM) and Knowledge Graphs (KG) into semantic representations of traditional methods. In order to improve semantic understanding in complex news texts, we use LLMs' powerful text understanding ability to generate news representations containing rich semantic information. In addition, our method combines the information about news entities and mines high-order structural information through multiple hops in KG, thus alleviating the challenge of long tail distribution. Experimental results demonstrate that compared with various traditional models, the framework significantly improves the recommendation effect. The successful integration of LLM and KG in our framework has established a feasible path for achieving more accurate personalized recommendations in the news field. Our code is available at https://github.com/Xuan-ZW/LKPNR.

\end{abstract}

\section{Introduction}
With the exponential growth of the Internet, an increasing number of individuals are opting to access the most current global news via online platforms, including MSN News. However, users often find themselves overwhelmed by the sheer volume of available news content. Consequently, the imperative for an effective news recommendation system becomes evident, as it serves as a crucial tool in assisting users to navigate through this vast information landscape. Such a system not only aids users in filtering copious amounts of news but also employs personalized algorithms to proactively present news items aligning with users' genuine interests, thereby significantly enhancing the fulfillment of their informational requirements\cite{Wu2021PersonalizedNR}.

\begin{figure}
    
         \includegraphics[width=0.5\textwidth]{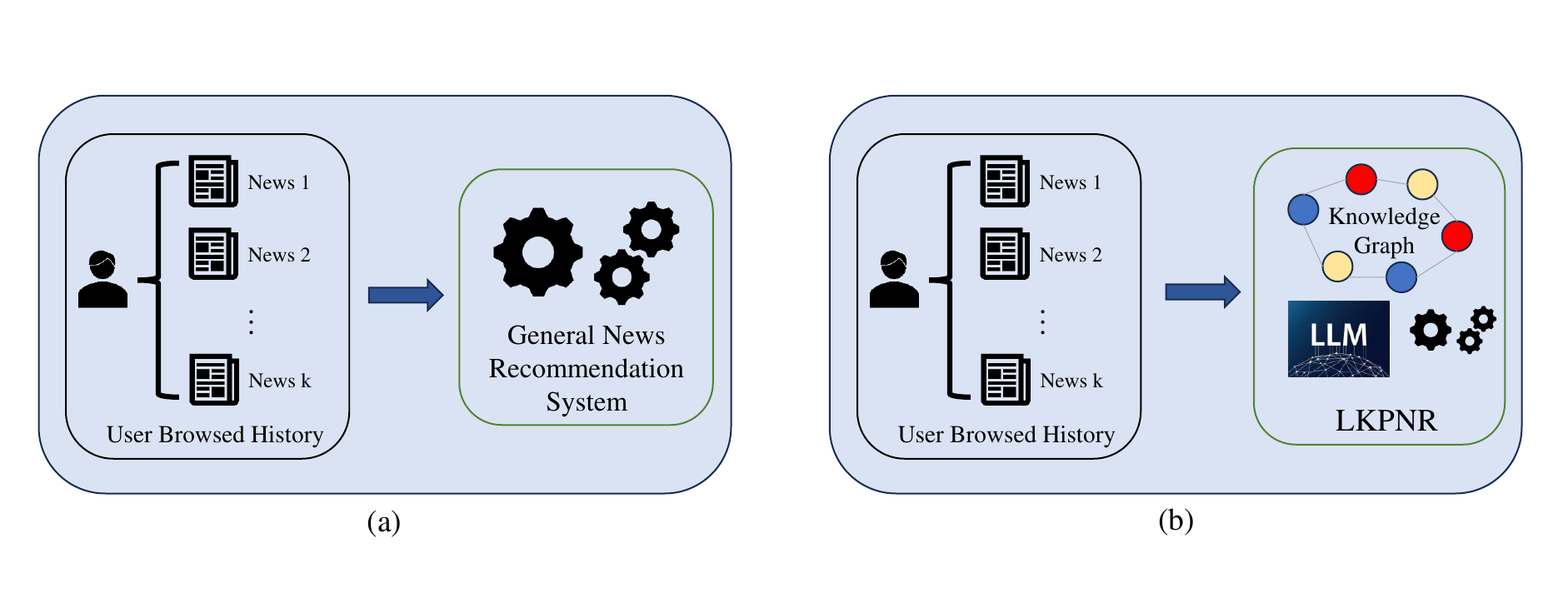} 
         \caption{Differences between traditional news recommendation model (a) and LKPNR framework (b)}
\end{figure}
The current research in this field primarily emphasizes the perspective of representation learning\cite{Bengio2012RepresentationLA}, aiming to enhance the learning of user and news representations separately. To illustrate this, we take the MIND dataset\cite{Wu2020MINDAL} as a case study. The user's behavioral data within this dataset comprises a sequence of historical clicks on news items, each of which is composed of title, abstract, category and other related information. Earlier researches, exemplified by NAML\cite{Wu2019NeuralNR} and NRMS\cite{Wu2019NeuralNR1}, have been employed to achieve improved news representations. The methods often leverage CNN and LSTM for feature extraction, coupled with feature fusion using attention mechanisms. However, these existing studies exhibit certain limitations that deserve attention. One such problem is insufficient news text feature extraction, this deficiency hampers the effectiveness of the news representation process. Neglect of news popularity and interconnections is another challenge, this neglect can potentially lead the news recommender system into the challenging scenario of the long-tail problem.

With the advent of ChatGPT, the realm of natural language processing has seamlessly ushered in the era of Large Language models (LLMs). Notably, significant models like ChatGLM2\cite{Zeng2022GLM130BAO}, LLAMA2\cite{Touvron2023LLaMAOA}, and RWKV\cite{Peng2023RWKVRR} have surfaced within the open-source community. LLMs, having undergone pre-training on vast corpora of textual data, exhibit the ability to swiftly acclimate to the data distribution pertaining to downstream tasks. Leveraging the exceptional language modeling proficiency of LLMs, they adeptly uncover intricate linguistic relationships and semantic nuances inherent within the text. This capacity allows for a more robust contextual integration, thereby augmenting text comprehension and facilitating the extraction of information-rich semantic features.

The long-tail problem\cite{Yin2012ChallengingTL} in recommendation systems is that a significant majority, approximately 80\%, of user clicks are concentrated on a mere 20\% of popular items. Consequently, this tendency results in recommendation systems favoring these popular items, often overlooking less popular ones, which, over time, detrimentally impacts the overall effectiveness of recommendations.  To address this long-tail challenge, recent research\cite{Guo2020ASO, Wang2019KnowledgeGC} has explored the incorporation of knowledge graphs (KGs) as supplementary information for recommender systems. This innovative approach leverages graph-based learning\cite{Ouyang2021LearningGM} to establish meaningful relationships among diverse items within the system. Subsequently, this method harnesses additional item-specific information to enhance the representation of long-tailed items, effectively mitigating the issue of inadequate representation learning for such items.

To address the aforementioned challenges, we introduce LKPNR (\underline{L}LM and \underline{K}G for \underline{P}ersonalized \underline{N}ews \underline{R}ecommendation framework), a personalized news recommendation framework that links general news recommendation models with the integration of KGs and LLMs. Capitalizing on the robust text comprehension abilities of the LLM, we generate news representations imbued with comprehensive semantic information for each news item, thereby mitigating the shortcomings associated with the limited feature extraction capabilities inherent in general news recommendation models. Concurrently, the incorporation of subgraph structural representations, mined through multi-hop inference within KG, serves to alleviate the issue of long-tail distribution prevalent in news recommendation. By harnessing the strengths of both LLM and KG, we observe a substantial enhancement in the model's performance. The differences between our proposed framework and the general news recommendation model are shown in Figure 1. To summarize, our contributions are listed as follows:

\begin{itemize}
	\item We propose LKPNR, a personalized news recommendation framework that fuses general news recommendation models with KG and LLM. To the best of our knowledge, this is the first work that combines both KG and LLM in the news recommendation domain.
	\item LKPNR can be flexibly combined with various general news recommendation models. Leveraging LLM$'$s powerful text understanding and the news graph structural relationships contained in the KG to inject additional information into general news recommendation models.
	\item Experiments on the MIND dataset show that LKPNR can significantly improves the performance of general news recommendation models.
\end{itemize}

\section{Related Work}
\subsection{General News Recommendation Model}
General news recommendation models typically involve encoding various aspects of news, such as title and abstract independently. These encoded representations are then interacted with separately to create a comprehensive news representation. In a similar manner, historical news browsing sequences are encoded and integrated to form a user representation. Subsequently, the similarity between this user representation and the representations of candidate news items is computed, enabling the prediction of whether the user would find these candidates interested or not. Okura et al.\cite{Okura2017EmbeddingbasedNR}extract news features through a denoising self-encoder, while user sequences are derived using RNN to obtain user features. Lian et al. proposed DFM\cite{Lian2018TowardsBR}using multi-channel inception blocks and attention mechanism to tackle the issue of data diversity. An et al.\cite{An2019NeuralNR} employed GRU to model both long-term and short-term interests, resulting in improved user representations. Wu et al. extended the attention mechanism\cite{Vaswani2017AttentionIA} paradigm in several ways\cite{Wu2019NeuralNR, Wu2019NPANN}. Wang et al.\cite{Wang2020FinegrainedIM} employ fine-grained interest matching using dilated convolution and 3D convolution techniques.

\subsection{LLM-Powered News Recommendation Model}
With the remarkable performance exhibited by LLM across diverse domains, researchers have embarked on an exploration of its potential within the recommendation domain. Kang et al.\cite{Kang2023DoLU} conducted a comparative study encompassing traditional collaborative filtering methods and LLM for user rating prediction, examining zero-shot, few-shot, and fine-tuned scenarios. Their research revealed intriguing insights. Likewise, Hou et al.\cite{Hou2023LargeLM} devise various prompts to address the ranking predicament in recommender systems. Gao et al.\cite{Gao2023ChatRECTI} transform user profiles and historical interactions into prompts, leveraging ChatGPT$'$s in-context learning capabilities for recommendation.

The above methods utilize LLM's in-context learning capabilities by constructing prompts to cope with downstream tasks in the general recommendation domain, but the performance is far inferior to that of traditional ID embedding-based fine-tuning methods. In the news recommendation domain, LLM is also beginning to combine with traditional models. Liu et al.\cite{Liu2023AFL} use ChatGPT to generate user profiles and augment the news text, combining with a traditional recommendation model to achieve better results. Li et al.\cite{Li2023ExploringTU} based on the traditional recommendation model, generated news representations by using LLM as a news encoder directly to complete the news recommendation.

\begin{figure*}
	\centering		
         \includegraphics[width=1.0\textwidth]{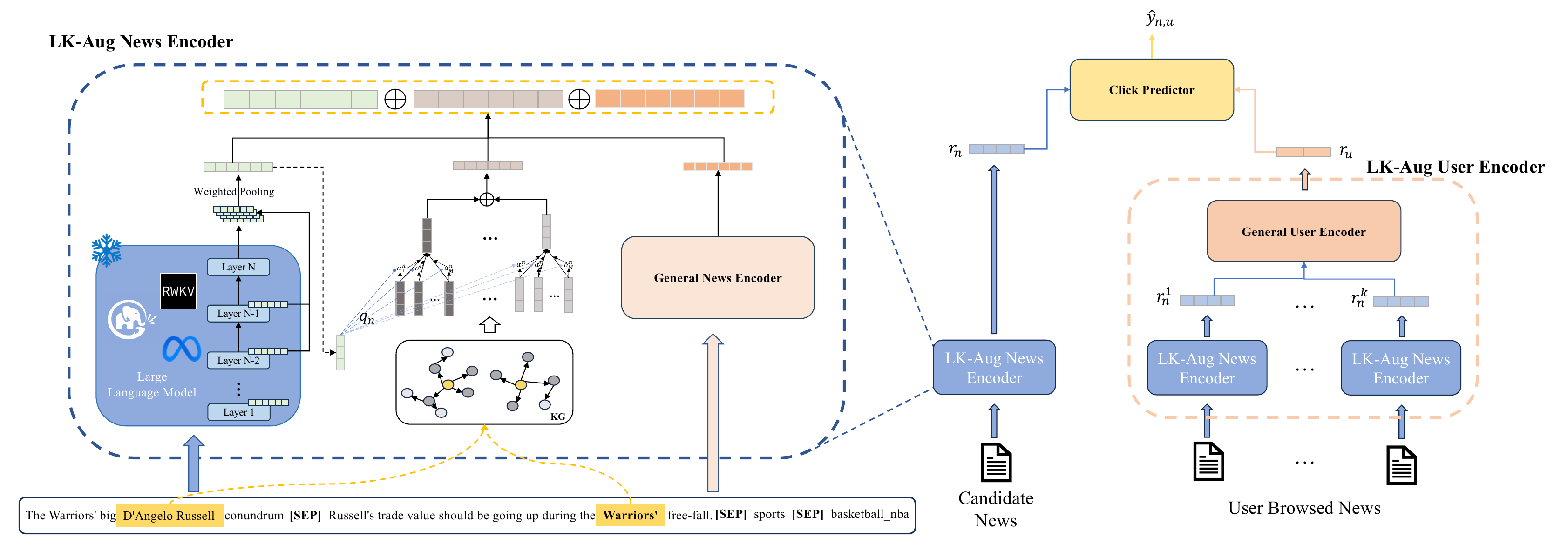} 
         \caption{The framework of LKPNR. The lower left corner of figure is the input news data into LKPNR, where the words marked in yellow indicates the entities. Candidate news is encoded by LK-Aug news encoder that combines LLM and KG with traditional general news encoder to obtain news representation. To obtain user representation, LK-Aug User Encoder contains several LK-Aug News Encoders, which encodes user's historical click behaviors.
}
\end{figure*}
\subsection{KG-powered News Recommendation Model}
Recommendation systems predominantly operate within a graph structure due to the information-rich nature of the data, which leads to the widespread adoption of GNN within the realm of news recommendation. Wang et al.\cite{Wang2018DKNDK} proposed DKN, which constructs a graph of entities within the input news and was utilized to seek neighboring nodes for expanding news information. Another significant contribution by Mao et al\cite{Mao2021NeuralNR}. involves the integration of a GCN with the user's browsing history and leverages both LSTM and attention mechanisms to extract textual information. Yang et al.\cite{Yang2023GoingBL} employ a gated GNN, combining global representations from other users with local representations to enhance the personalized recommender system. Furthermore, several research endeavors aim to amplify representation and inference capabilities by linking LLM and KG. Sun et al.\cite{Sun2021JointLKJR} utilized attention mechanism to facilitate interaction between all tokens of the input text and KG entities, thus augmenting the output of the language model with domain-specific knowledge from the KG.

It is evident that the KG within the news domain holds substantial value in terms of structured information and domain-specific knowledge. Its inherent incompleteness and limited language understanding are effectively supplemented by the extensive general knowledge encapsulated within LLM. Notably, there is currently a dearth of work that combines LLM and KG within the news recommendation field. Therefore, we propose LKPNR as a solution to bridge this gap.

\section{Problem Formulation}\

The personalized news recommendation system proposed takes the user's news clicks and the KG of the entities within the news as input. The user's news clicks refer to a number of pairs of user and news, denoted as $\mathcal{C}=(u,n) \subseteq \mathcal{U} \times \mathcal{N}$, where $\mathcal{U}$ is the user set and $\mathcal{N}$ is the news set. The KG of the entities contained in the news set $\mathcal{N}$ is a triple $\mathcal{G}=(h,r,t) \subseteq \mathcal{E} \times \mathcal{R} \times \mathcal{E}$, where $h,t \in \mathcal{E}$ represent the head and tail entities, and $r\in \mathcal{R}$ represents the relation between them. Given the training click set $\mathcal{H}$ and test click set $\mathcal{T}$, where $\mathcal{H} \cup \mathcal{T}=\mathcal{C}$ and $\mathcal{H}\cap\mathcal{T}=\phi$. Then the recommendation task can be formulated as learning a matching function $\mathcal{F}$ from the training set $\mathcal{H}$ and predicting the degree of user-news matching in the test set $\mathcal{T}$ to obtain the score $\mathcal{M}$.

\section{Framework}
The overall recommendation system framework consists of three components: the LLM and KG Augmented News Encoder (LK-Aug News Encoder), the LLM and KG Augmented User Encoder (LK-Aug User Encoder), and the Click Predictor. The overall recommendation framework is illustrated in Figure 2

\subsection{LK-Aug News Encoder}
The LK-Aug News Encoder is composed of three sub-modules: General News Encoder, LLM-Augmented Encoder and KG-Augmented Encoder.

\subsubsection{General News Encoder.} 
The general news encoder is designed to learn a semantic representation of news at the word level, which is achieved through a structure that typically consists of two layers: the word embedding layer and the word fusion layer. The word embedding layer utilizes an embedding matrix $ E \in R^ {(|v| \times d)}$ to convert each word $w$ occurring in the news title and abstract into an embedding vector $e$. Here, $v$ denotes the number of words and $d$ denotes the dimension of the word embedding. The fusion layer is a well-designed component in the baseline experiments that interacts with the embedding vectors of individual words via operations such as linear mapping, weighted summing, and vector concatenation. These operations fuse the vectors to produce a generic representation vector $r_{GNE}$ for the given news.


\subsubsection{LLM-Aug Encoder.} 
We leverage powerful contextual understanding capability of LLM and rich open-world knowledge to build news representations with comprehensive semantic information for the input news to solve the problem of limited text feature extraction capability in general news recommendation. The input news text is created by concatenating news titles, news abstract, category and sub-category through [SEP], as shown the Figure 2.

\begin{equation}\label{eq1}
S_{HS}=LLM(t)
\end{equation}
\begin{equation}\label{eq2}
S_{HS\_p}= mean\_pooling(S_{HS})   
\end{equation}
where $LLM(\cdot)$ is the LLM decoder which returns the hidden states of the last four decoder layers, denoted $S_{HS}$, and $S_{HS_p}$ denotes the output after taking mean pooling strategy on the sequence length dimension of the each layer.
After that $S_{HS_p}$ will be weighted and summed to get $S_W$
\begin{equation}
S_W=\sum\limits_{i=1}^{4}(a_iS_{HS\_p}^i)    
\end{equation}
where $a_i$ denotes the learnable weights, $S_{HS\_p}^i$ denotes the hidden states of the $i-th$ layer.
Finally the weighted hidden state $S_W$ will be projected to the text representation space by nonlinear mapping. 
\begin{equation}
    r_{LLM}=\sigma(f_l(S_W))
\end{equation}

where $\sigma$ denotes the activation function, $f_l$ denotes the linear transformation, and $r_{LLM}$ is the enhanced news representation.

\subsubsection{KG-Aug Encoder}
We feed the enhanced news representation $r_{LLM}$ into the nonlinear layer, which is used to transform $r_{LLM}$ from the textual representation space to the entity representation space.

\begin{equation}\label{eq1}
q=\sigma(f_s(r_{LLM}))
\end{equation}
where $f_s$ is a linear transformation. The query $q$ will have extensive interactions with the entities in the KG.

Generally, the title and abstract of a piece of news will contain several source entities. Considering the 1,2,…,n hop adjacent entities of these source entities, we can extract a subgraph $g^n=(V, R)$ from the external KG, where $V$ is the set of entities in the subgraph, $R$ is the set of edges connecting entities. Taking the $k-th$ hop adjacent entity set as an example, $V^k=\{v_i^k\}_{i=0}^{|M|}$ represents $M$ entities with $k$ hops from the source entity. Then we get the embedding vector corresponding to each entity through the wiki KG, denoted as $X^k=\{x_i^k\}_{i=0}^{|M|}$. Given a query $q$ and a neighboring entity set $X^k$ with hop count $k$, the KG-Augmented Encoder interacts the query with each vector $x_i^k$ in $X^k$ to generate a news attention score for the entity, denoted as $\alpha_i^k$.
\begin{equation}
\alpha_i^k=Softmax(W^T_k[q;x_i^k;q\circ x_i^k])    
\end{equation}
where $W_{\alpha k}^T\in R^{3l\times 1}$, is a learnable parameter matrix, $\circ$ is the element multiplication, and $[;]$ denotes the vector connection.

The weighted representation $\hat{x}^k$ of a $k$-hop entity can be computed as:
\begin{equation}
    \hat{x}^k=\sum_{i=1}^M{\alpha_i^kx_i^k}
\end{equation}

After that, we concatenate the weighted representation vectors of each hop and project the concatenated vectors to the news representation space, denoted as the KG representation $r_{KG}$

\begin{equation}
    r_{KG} = Q^T[\hat{x}^1;...;\hat{x}^n]
\end{equation}

where $Q^T\in R^{nl\times o}$, $nl$ denotes the dimension after the representation vector connection of each hop and $o$ represents the dimension of the projected KG representation.

The final news representation vector $r_n$ is obtained by connecting three news representations above.

\begin{equation}
    r_n =[r_{GNE};r_{KG};r_{LLM }]
\end{equation}

\subsection{LK-Aug User Encoder}

The LK-Aug User Encoder learns representations of users based on their click history on news. This module includes a News Embedding Layer (NEL) and a Representation Fusion Layer (RFL). The NEL obtains the representation of the news browsed by the user through the LK-Aug News Encoder, denoted as $[r_n^1,r_n^2,…,r_n^z  ]$, where $z$ represents the length of the historical browsing news sequence. The RFL transforms news representation sequences into user representations $r_u$ through a series of fusion methods, such as concatenation, mapping, attention, etc.

\begin{equation}
    [r_n ^ 1, r_n ^ 2,..., r_n ^ k]=NEL({h_1, h_2, ..., h_k})
\end{equation}

\begin{equation}
    r_u = RFL([r ^ 1, r ^ 2,..., r ^ k])
\end{equation}


\subsection{Click Predictor and Model Training}

Given candidate news and user representations $r_n$ and $r_u$, this module is used to get the matching score between user u and the candidate news n. We compute the dot-product$\hat{y}_{n,u}$  of $r_n$ and $r_u$ as the unnormalized matching scores of users and news.
\begin{equation}
    \hat{y}_{n,u}=\left \langle r_n, r_u \right \rangle
\end{equation}
Following previous general work, we use a negative sampling strategy for model training. For the $i-th$ news exposure, we compute its unnormalized matching score as $\hat{y}_i^+$. In addition, randomly select $K$ pieces of news as the news that the user does not click, and its unnormalized matching score is $[\hat{y}_{i,1}^-,\hat{y}_{i,2}^-,...,\hat{y}_{i,K}^-,]$. We employ softmax function to calculate the normalized matching score.
\begin{equation}
    p_i=\frac{exp(\hat{y}_i^+)}{exp(\hat{y}_i^+) + \sum_{j=1}^K{exp(\hat{y}_{i,j}^-)}}
\end{equation}

In this way, the click prediction problem can be formulated as a K+1 classification problem. For the training set $\mathbcal{H}$, the loss function $\mathbcal{L}$ of model training is the negative log likelihood of all positive samples, which can be described as follows:
\begin{equation}
    \mathbcal{L}=-\sum_{i\in \mathbcal{H}}log(p_i)
\end{equation}

\renewcommand{\arraystretch}{1.5} 
\begin{table*}[]
\begin{tabular}{|c|cccc|cccc|}
\hline
\multirow{3}{*}{Metric} &
  \multicolumn{4}{c|}{NRMS} &
  \multicolumn{4}{c|}{NAML} \\ \cline{2-9} 
 &
  \multicolumn{1}{c|}{\multirow{2}{*}{Orig}} &
  \multicolumn{3}{c|}{LKPNR} &
  \multicolumn{1}{c|}{\multirow{2}{*}{Orig}} &
  \multicolumn{3}{c|}{LKPNR} \\ \cline{3-5} \cline{7-9} 
 &
  \multicolumn{1}{c|}{} &
  \multicolumn{1}{c|}{LKPNR-NRMS} &
  \multicolumn{1}{c|}{w/o KG} &
  w/o LLM &
  \multicolumn{1}{c|}{} &
  \multicolumn{1}{c|}{LKPNR-NAML} &
  \multicolumn{1}{c|}{w/o KG} &
  w/o LLM \\ \hline
AUC &
  \multicolumn{1}{c|}{0.6802} &
  \multicolumn{1}{c|}{0.7049} &
  \multicolumn{1}{c|}{0.6997} &
  0.6845 &
  \multicolumn{1}{c|}{0.6842} &
  \multicolumn{1}{c|}{0.7023} &
  \multicolumn{1}{c|}{0.6995} &
  0.6841 \\ \hline
MRR &
  \multicolumn{1}{c|}{0.3316} &
  \multicolumn{1}{c|}{0.3492} &
  \multicolumn{1}{c|}{0.3441} &
  0.3308 &
  \multicolumn{1}{c|}{0.3257} &
  \multicolumn{1}{c|}{0.3423} &
  \multicolumn{1}{c|}{0.3411} &
  0.3319 \\ \hline
nDCG@5 &
  \multicolumn{1}{c|}{0.3661} &
  \multicolumn{1}{c|}{0.3886} &
  \multicolumn{1}{c|}{0.3846} &
  0.3680 &
  \multicolumn{1}{c|}{0.3621} &
  \multicolumn{1}{c|}{0.3816} &
  \multicolumn{1}{c|}{0.3810} &
  0.3699 \\ \hline
nDCG@10 &
  \multicolumn{1}{c|}{0.4306} &
  \multicolumn{1}{c|}{0.4514} &
  \multicolumn{1}{c|}{0.4453} &
  0.4310 &
  \multicolumn{1}{c|}{0.4257} &
  \multicolumn{1}{c|}{0.4440} &
  \multicolumn{1}{c|}{0.4426} &
  0.4325 \\ \hline
\end{tabular}
\caption{Performance of comparison results (Orig. denotes general news recommendation, LKPNR-NRMS/NAML denotes NRMS/NAML with LKPNR, w/o KG denotes remove the KG-Augmented Encoder, w/o LLM denotes remove the LLM-Augmented Encoder)}
\end{table*}

\section{Experiments}
\subsection{Dataset}
We conduct experiments on the MIND dataset, which is a dataset constructed based on user click logs on the MSN online news site. We use the same processing method as Mao et al. We randomly sample 200K user click logs from the training and validation sets of the MIND dataset. Given the absence of labels for the test set, we partition the original validation set into two distinct segments: the experimental validation set and the experimental test set. The specific information of the constructed sampled dataset is shown in Table 2.

\renewcommand{\arraystretch}{1.5} 
\begin{table}[t]

\begin{tabular}{cc|cc}
\hline
\# users & 20000 & \#user in train set & 189580  \\
\hline
\# news & 78602 & \# news in train set & 75963 \\
\hline
\# training logs & 595186 & \# training samples & 905297 \\
\hline
\end{tabular}
\caption{Statistics of MIND-200K dataset}
\end{table}

\subsection{Implementation Details and Metrics}
We select NAML and NRMS as our baseline models, setting the maximum sequence length for user browsing history to 50 and adopting a negative sampling rate of $k=4$. We maintain the parameter configurations of the General News Encoder identical to the baseline. In the LLM-augmented encoder, the hidden states of the LLM are projected to 500 dimensions. For the KG-augmented encoder, the dimensions of entity embeddings are set to 100. Furthermore, we limit the maximum number of neighboring nodes to 20 per source node, and the traversal depth is constrained to a maximum of 2 hops. Throughout the training process, learning rate employs 1e-4, batch size is set to 64, and early-stop strategy is implemented. All experiments are conducted on the NVIDIA TESLA V100.

To evaluate the model's performance, we use four widely recognized evaluation metrics, specifically, the Area Under the ROC Curve (AUC), Mean Reciprocal Rank (MRR), and Normalized Discounted Cumulative Gain (nDCG@5 and nDCG@10).

\subsection{Performance Comparison}
We perform LKPNR on two baseline experiments. To further validate the efficacy of our framework design, we conduct ablation experiments, the experimental results are summarized in Table 1.

The experimental results show that the two baseline models have significant improvement in all evaluation metrics (+2.47\%/1.81\% AUC, +1.76\%/1.66\% MRR, +2.25\%/1.95\% nDCG@5, and +2.08\%/1.83\% nDCG@10 compared to NRMS/NAML performance) with the augmentation of our framework. This performance improvement derives from the fact that the news encoder of the baseline gains better performance through the enhanced semantic information of the LLM and the collaborative information of the KG. As depicted in Table 1, discernible decrements in performance are across the spectrum of ablation variants compared to our complete model, which demonstrates the efficacy of the different components of our model. The removal of the LLM exhibits a substantial impact on the overall performance of the model, demonstrating the effectiveness of the augmenting semantic information for news representation.

\subsection{Performance of Different LLM}
The characteristics of LLM can vary due to inconsistencies in the proportion of data categories within the training data and model structural design. Consequently, these differences lead to varying capabilities among LLMs, reflected in their open-world knowledge, performance on diverse tasks and so on, which leads to their distinct understanding of text. For instance, ChatGLM2\cite{Zeng2022GLM130BAO} is trained on the same amount of Chinese and English corpus, and can handle Chinese and English tasks with high quality. LLAMA2\cite{Touvron2023LLaMAOA}, trained on an extensive, high-quality English dataset, demonstrates adeptness in handling various English tasks. RWKV\cite{Peng2023RWKVRR} exhibits a quicker reasoning speed and lower computational complexity. In order to explore the impact of various LLMs on news recommendation, we employ three cutting-edge models: ChatGLM2, LLAMA2, and RWKV to generate enhanced news representations on the basis of the baseline model NRMS. Table 3 shows the performance comparison of the enhanced news representation of different LLMs in the recommendation task.

The outcomes of our experiments utilizing these three diverse LLMs are detailed in the table below, and the results indicate that ChatGLM2 provides the most effective enhancement for news recommendation when compared to both LLAMA2 and RWKV. The reason for this phenomenon is that the training data of ChatGLM2 may contain a certain proportion of English news.

\renewcommand{\arraystretch}{1.5} 

\begin{table}[]
\begin{tabular}{|c|ccc|}
\hline
\multirow{2}{*}{Metrics} & \multicolumn{3}{c|}{LKPNR}                                                   \\ \cline{2-4} 
                         & \multicolumn{1}{c|}{ChatGLM2-6B} & \multicolumn{1}{c|}{LLAMA2-13B} & RWKV-7B \\ \hline
AUC                      & \multicolumn{1}{c|}{0.7049}      & \multicolumn{1}{c|}{0.6845}     & 0.6771  \\ \hline
MRR                      & \multicolumn{1}{c|}{0.3492}      & \multicolumn{1}{c|}{0.3307}     & 0.3300  \\ \hline
nDCG@5                   & \multicolumn{1}{c|}{0.3886}      & \multicolumn{1}{c|}{0.3657}     & 0.3631  \\ \hline
nDCG@10                  & \multicolumn{1}{c|}{0.4514}      & \multicolumn{1}{c|}{0.4307}     & 0.4218  \\ \hline
\end{tabular}
\caption{the performance comparison of the enhanced news representation of different LLMs}
\end{table}


\subsection{Effectiveness of KG Entity's Query}
We employ comparative experiments to ascertain optimal strategies for the retrieval and integration of adjacent entities in a more efficient manner. The incorporation of neighboring entity vectors in the news coding process can be perceived as a mechanism that augments the collaborative information. By fusing vectors of neighboring entities, the KG-Augmented Encoder is able to gather information from multiple related entities and synergistically integrate them into a single encoded representation. The crux of enhancing news representation through the use of KG lies in the efficient extraction of information from all neighboring entities. In KG-Augmented Encoder, the query is obtained by projecting the textual representation of news into the entity representation space. We have implemented an experiments to compare the retrieval performance of the query converted by general news encoder and query converted by LLM.
In addition, direct weighted summation of all entities at each hop may lead to relatively large information loss. We use multi-head query to consider different representation spaces, i.e., the multiple weighted summation of all entities with different weights can capture collaborative information from different perspectives of the news and entities, and thus improves the representational capability of KG-Augmented Encoder. The detailed experimental results are shown in Figure 3.
\begin{figure}
	\centering		
         \includegraphics[width=0.5\textwidth]{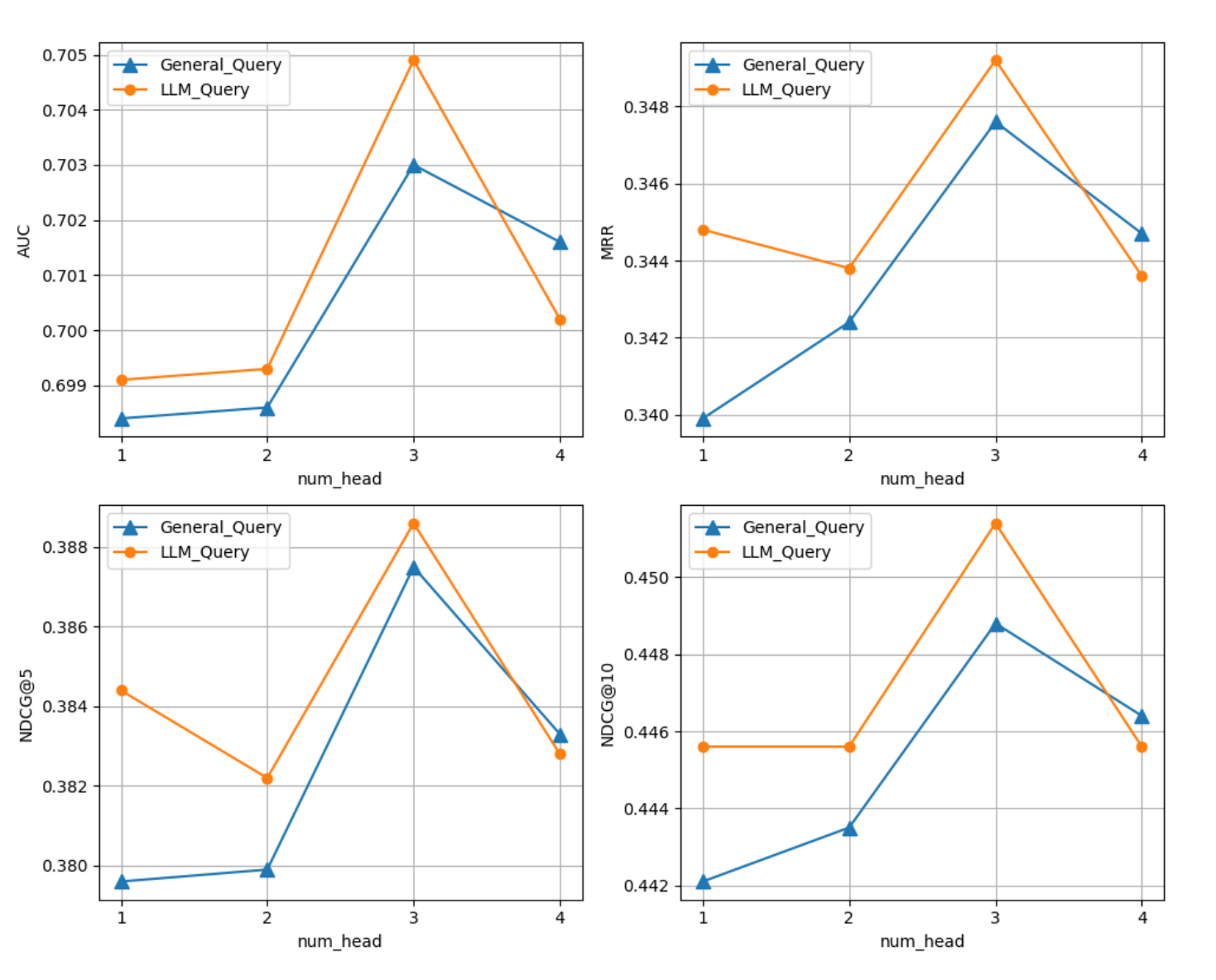} 
         \caption{Comparison of different queries under different num heads(General\_query denotes General News Encoder mapping query, LLM\_query denotes LLM mapping query)}
\end{figure}

The experimental results show that the retrieval performance of LLM mapping query outperforms the General News Encoder mapping query. Compared with the news representation produced by the General News Encoder, which is limited to the specifics of the news text, the news representation of the LLMs contains some open world knowledge about the news, and is thus able to understand the information of neighboring entities more effectively. In addition, when num\_head=3, the LLM demonstrates its highest proficiency in mapping queries to extract information. It suggests that expanding the representation space of the query vector enhances its capability to gather entity information.

\section{Case Study}

\begin{figure*}
	\centering		
         \includegraphics[width=0.8\textwidth]{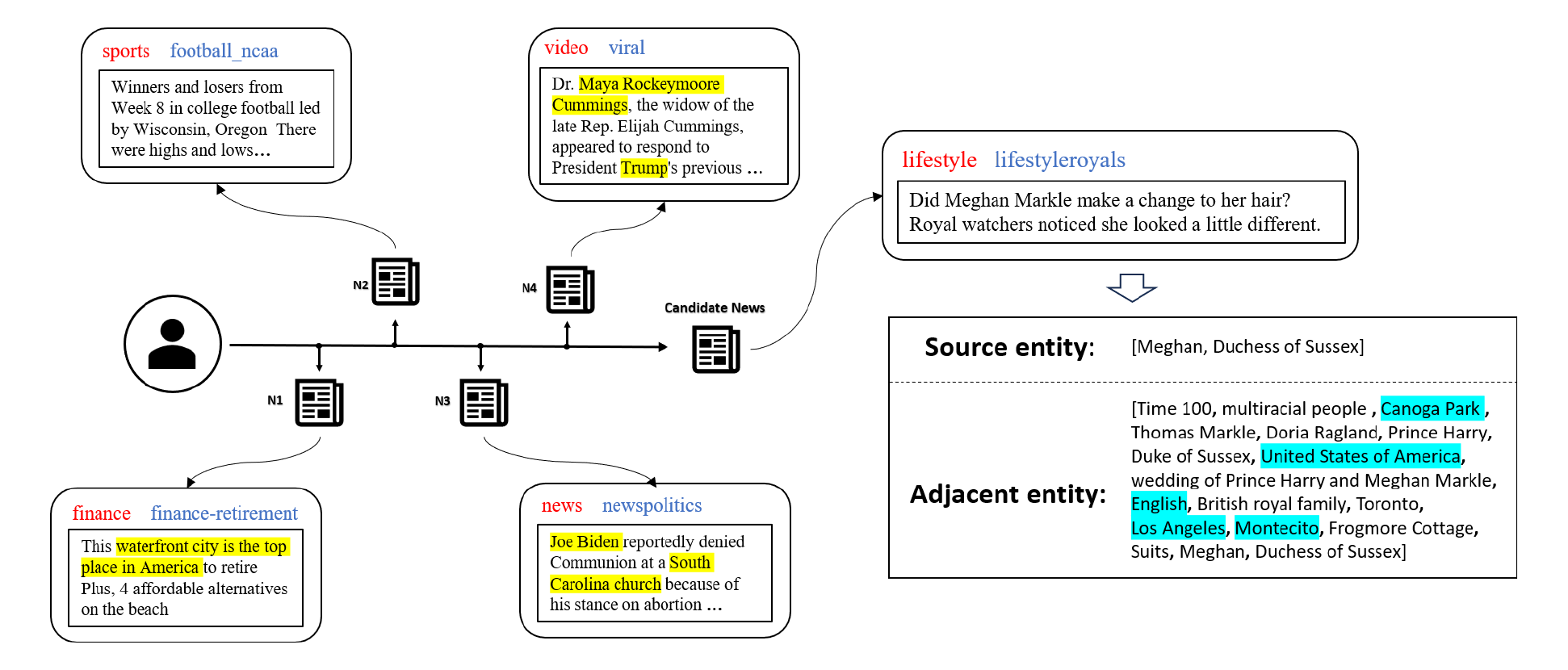} 
         \caption{Historical click sequence and candidate news for the sample user}
\end{figure*}

\begin{figure}
	\centering		
         \includegraphics[width=0.5\textwidth]{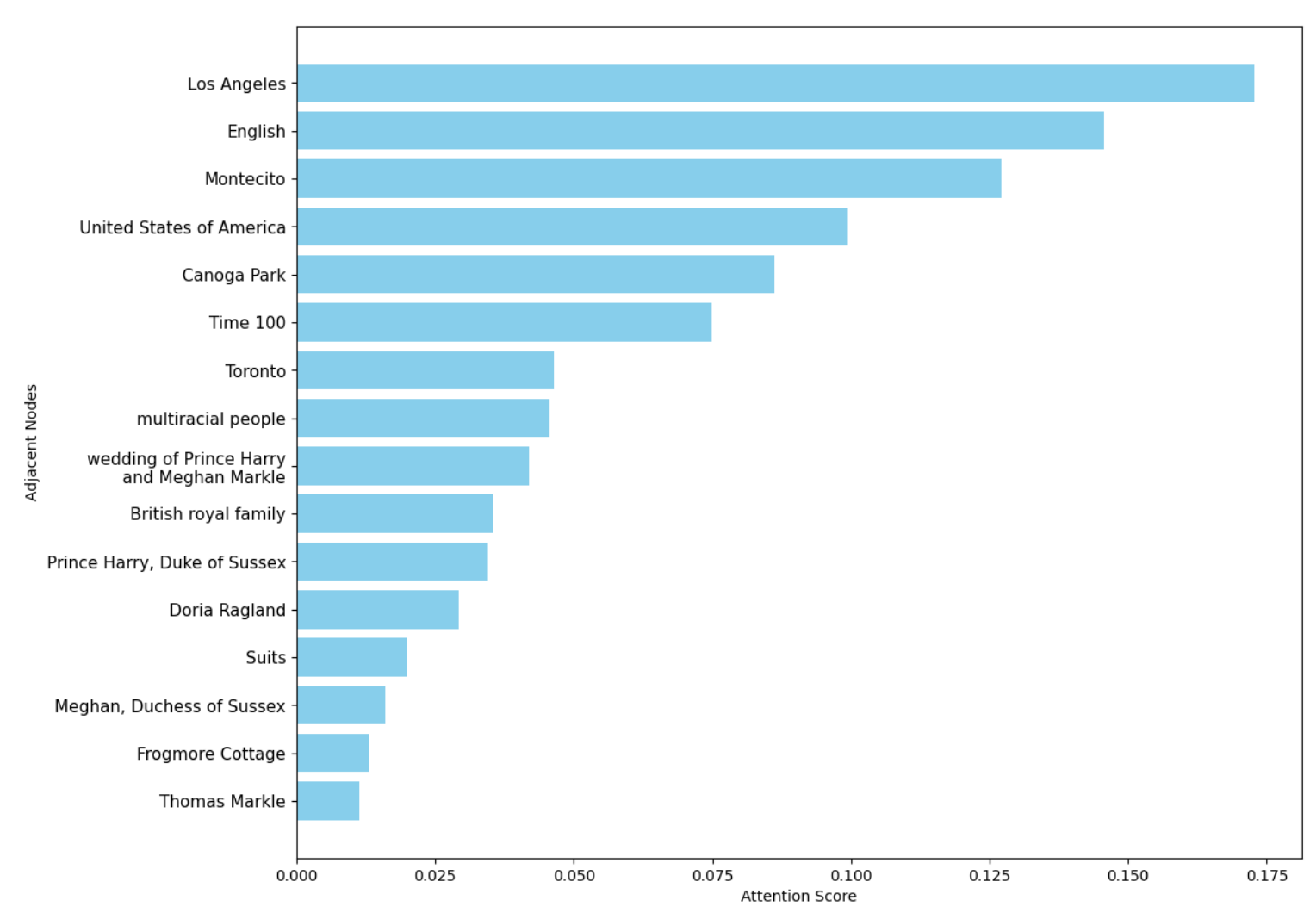} 
         \caption{The visualization of the attention weights to adjacent nodes}
\end{figure}
In this section, we demonstrate the characteristics of LLM and KG Augmented News Encoder using visualization. Figure 4 shows a user's click order of historical news and current news candidates, as well as the detailed categories, subcategories, titles and abstracts of these news. Before the integration of KG, the user's expression was transformed from the text and category characteristics of these historical news clicks. This candidate news has little correlation with any historical news clicked by users, and the matching degree between the candidate news and the user vector is also very low. After KG integration, give full consideration to adjacent nodes. Figure 4 contains the source and neighboring entities of the candidate news, and the entities in the historical click news are highlighted in yellow. Figure 5 shows the attention of a query to neighboring nodes, and in Figure 4, the five nodes with the highest attention are highlighted in blue. Although this candidate news has a relatively low correlation with the user's historical click news, it has a large number of adjacent nodes with a high correlation with the historical click news. The historical news sequence contains a number of U.S. locations and celebrities, and the candidate news also includes many of these entities in its adjacent nodes. This shows that the candidate news has a deep potential connection with the historical click news, and the matching degree has been greatly improved after considering the neighbor nodes. 

LLM and KG Augmented News Encoder considers potential connections between candidate news and user's history of clicking on news, which makes user-candidate news matching beyond the understanding of news text.

\section{Conclusion}
In this work, we propose an innovative personalized news recommendation framework LKPNR, which integrates a Large Language Model (LLM) and a Knowledge Graph (KG). While combining the General News Encoder, the robust contextual comprehension capability of the LLM allows us to derive news representations imbued with semantic information. Simultaneously, we harness the news relationship graph structure inherent in the KG to extract supplementary collaborative news information, enhancing the efficacy of the news recommendation system and alleviating the long tail problem to a certain extent. The experimental results demonstrate the outstanding performance of our proposed framework, leading to significant enhancements over the traditional baseline.



\bigskip

\bibliography{aaai24}

\end{document}